\documentclass[
reprint,
amsmath,amssymb,
pre,
floatfix,
]{revtex4-1}

\usepackage{graphicx}% Include figure files
\usepackage{dcolumn}% Align table columns on decimal point
\usepackage{bm}% bold math
\usepackage[normalem]{ulem} % for striking through the command "\sout{...}"
\usepackage{titlesec}
\usepackage{dcolumn}% Align table columns on decimal point
\usepackage{bm}% bold math
\usepackage{graphicx}% Include figure files
\usepackage{afterpage}
\usepackage{xcolor}
\usepackage{amsmath}
\usepackage{amssymb}
\usepackage{graphicx}% Include figure files
\usepackage{bm}% bold math
\usepackage{epstopdf}
\usepackage{hyperref}
\usepackage{csquotes}
\usepackage{orcidlink}
\usepackage{booktabs, array}
\hypersetup{
	colorlinks=true,
	linkcolor=blue,
	filecolor=magenta,      
	urlcolor=blue,
	citecolor=red,
}

\newcommand{\Jav}{J_\mathrm{av}}

\begin{document}

\preprint{distribution symmetry}

\title{Effect of diversity distribution symmetry on global oscillations of networks of excitable units}% Force line breaks with \\
%\thanks{A footnote to the article title}%

\author{S.~Scialla$^{1,2}$\orcidlink{0000-0003-1582-8743}, M.~Patriarca$^1$\orcidlink{0000-0001-6743-2914}, E.~Heinsalu$^1$, M.~E.~Yamakou$^{3}$\orcidlink{0000-0002-2809-1739}, Julyan~H.~E.~Cartwright$^{4,5}$\orcidlink{0000-0001-7392-0957}}

\affiliation{$^1$  National Institute of Chemical Physics and Biophysics - R{\"a}vala 10, Tallinn 15042, Estonia; \\
$^2$ Department of Science and Technology for Sustainable Development and One Health, Università Campus Bio-Medico di Roma - Via Á. del Portillo 21, 00128 Rome, Italy; \\
$^3$ Department of Data Science, Friedrich-Alexander-Universit\"{a}t Erlangen-N\"{u}rnberg, Cauerstr. 11, 91058 Erlangen, Germany; \\
$^4$ Instituto Andaluz de Ciencias de la Tierra, CSIC, 18100 Armilla, Spain; \\
$^5$ Instituto Carlos I de Física Teórica y Computacional, Universidad de Granada, 18071 Granada, Spain}

\vspace{0.5cm}

\date{\today}% It is always \today, today

\begin{abstract}
We investigate the role of the degree of symmetry of the diversity distribution in shaping the collective dynamics of networks of coupled excitable units modeled by FitzHugh–Nagumo equations. While previous studies have focused primarily on the ratio between the numbers of individually oscillatory and excitable units, we show that the symmetry of the diversity distribution plays a fundamental role in the emergence of global network oscillations. 
By exploring various symmetric and asymmetric distributions and simulating network dynamics across various topologies, we demonstrate that symmetric distributions promote resonant collective oscillations even in the absence of oscillatory units. We propose two quantitative metrics, the normalized center of mass and the symmetry balance score, to assess the degree of symmetry and predict the presence or absence of global oscillations.
By studying a minimal two-unit system and its effective pseudo-potential, we show that symmetry enables the formation of a landscape characterized by a cyclic valley supporting limit cycles, whereas asymmetry collapses the system into a single non-oscillatory equilibrium. These results provide a general mechanism by which network symmetry drives emergent synchronization in heterogeneous excitable systems.
\end{abstract}

%\keywords{Suggested keywords}%Use showkeys class option if keyword display desired

\maketitle

%\tableofcontents

%%%%%%%%%%%%%%%%%%%%%%%%%%%%%%%%%%%%%%%%%%%%%%%%%%%%%%%%%%%%%%%%%%%%%%%%
\section{INTRODUCTION}
\label{Sec_Intro}
%%%%%%%%%%%%%%%%%%%%%%%%%%%%%%%%%%%%%%%%%%%%%%%%%%%%%%%%%%%%%%%%%%%%%%%%5

The impact of diversity, or heterogeneity, on the global oscillations of a network of coupled excitable units has been the subject of many studies since the last decade of the 20th century~\cite{Cartwright-2000a,Tessone-2006a,Toral-2009a,Chen-2009a,Wu-2010a,Wu-2010b,Patriarca-2012a,Tessone-2013a,Grace-2014a,Patriarca-2015a,Liang-2020a,degliesposti-2001a,Li-2012a,Li-2014a,Gassel-2007a,Zhou-2001a,Glatt-2008a}. Beyond its intrinsic relevance to the theory of dynamical systems, the reason for the enormous amount of interest in these excitable media lies in the possibility to use them as models for complex and important biological oscillators, such as the heart and pancreas~\cite{BARRIO2020105275,Lebert2023,alonso2016nonlinear,wellner,FelixMartinez2023}.
A fundamental question related to the understanding of these systems concerns the mechanisms leading to collective oscillations or quiescence of a network of a combination of diverse units, some of which are individually in an excitable (non-oscillatory) state, while others are individually oscillatory.

Previous studies have focused on the ratio between the number of excitable and oscillatory units as a key parameter governing the emergence of global network oscillations or quiescence.
%In 2006, 
Tessone et al. showed that the occurrence of the so-called diversity-induced resonance (DIR) in a network of FitzHugh-Nagumo (FHN) units, with linear all-to-all coupling, essentially depends on the ratio between the number of units in the excitable regime and those in an oscillatory state~\cite{Tessone-2006a}. 
In the system they studied, this ratio was determined by the degree of diversity applied to the population of FHN units and, at the same time, by an external forcing that had the effect of turning a certain fraction of units from excitable into oscillatory, by pushing them above their excitability threshold. The interpretation of the mechanism for the observed global network oscillations was that the units that became individually oscillatory were able, through the coupling terms, to ``pull" all the others producing collective oscillatory behavior.

%A few months before the publication of Tessone et al.~\cite{Tessone-2006a}, 
Paz\'{o} and Montbri\'{o} studied a population of Morris-Lecar units divided into two subgroups made up of identical excitable units and identical oscillatory units, respectively \cite{Pazo2006f}. 
The whole population was subjected to a linear all-to-all coupling. 
They also observed a pull effect between oscillatory and excitable units as the main mechanism driving global network oscillations. As a consequence, the collective behavior of the system could be predicted from the ratio of oscillatory to excitable units.

Analogous conclusions were drawn %two years later 
by Shen et al.~\cite{ShenChuan-Sheng_2008}, who studied a nearest-neighbor coupled, one-dimensional chain of heterogeneous FHN elements, to which they added random long-range connections. %; ``shortcuts''. 
Most or all units in their system were below the excitation threshold and could become oscillatory due either to an increase of diversity or to the effect of an external forcing. 
They found that collective network oscillations could be maximized by an optimal degree of diversity, combined with an ideal coupling strength and a higher number of long-range connections. 
Again, their mechanistic interpretation was that collective network oscillations were driven by individually oscillatory units being able to pull the excitable ones.

Global network oscillations can be observed even in heterogeneous systems that do not have any oscillatory units at all. 
%Some years prior to the above studies, %at the end of the 20th century 
Cartwright showed that a medium comprising diverse FHN units, which individually were all in an excitable, non-oscillatory state, was capable of spontaneously giving rise to collective network oscillations \cite{Cartwright-2000a}. 
The units formed a 3D network with cubic lattice topology, where each element was linearly coupled to six neighbors. 
In that work, diversity was introduced by dividing the network population into two subgroups of FHN units, one of which was made of identical elements that were all below the excitation threshold, while the other was made of identical elements that were all above the excitation block threshold. 
In other words, the two subgroups were symmetrically positioned, respectively, below and above the range of parameter values corresponding to the oscillatory regime in the FHN equations.

A wide range of systems that present analogies with the above-mentioned ones have been studied in the context of network aging processes \cite{Daido2004,Sharma2025,Morino,SHARMA2022,Liu_2016,Kundu,Rakshit2020,Sharma2021,Yuan2017}, where the coupling between varying fractions of spatially distributed active and inactive units determines the overall network activity. 
This is relevant to the modeling of biological systems where aging processes and/or the occurrence of a disease may cause a progressive increase in inactive cells, and it is therefore useful to determine the critical fraction of inactive cells above which the network loses its functionality.

In the present work, we dig deeper into the effect on global network oscillations of the degree of symmetry of the diversity distribution of a network of excitable units, modeled by the FHN equations as a prototypical example. 
We demonstrate that, in addition to the ratio between the number of individually oscillatory and excitable elements---which can only be applied to some specific network configurations---another, more general criterion determining the occurrence of global network oscillations or quiescence is linked to the degree of symmetry of the diversity distribution. 
This also implies that there are other mechanisms driving collective network oscillations beyond the simple pull of excitable units by those in an oscillatory state. 
We will show that the effect of diversity distribution symmetry can be observed across different network topologies.
Asymmetry may have physiological relevance, since the deficit in functional cell mass in a given organ that can derive from aging or disease is unlikely to equally affect all cell subgroups~\cite{vogt2016diversity,Efrat2019,Leenders,WAGNER2020136}. 
In modeling terms, this translates into a loss of symmetry of the diversity distribution that describes the cell ensemble.

%%%%%%%%%%%%%%%%%%%%%%%%%%%%%%%%%%%%%%%%%%%%%%%%%%%%%%%%%%%%%%%%%%%%%%%%5
\section{MODEL}
\label{sec_model}
%%%%%%%%%%%%%%%%%%%%%%%%%%%%%%%%%%%%%%%%%%%%%%%%%%%%%%%%%%%%%%%%%%%%%%%%%

%\subsection{FitzHugh-Nagumo model} 
%\label{sec_FN}

As a paradigmatic example of excitable systems, we study networks of FHN units with different topologies, i.e., cubic lattice, all-to-all coupling, and small-world (Newman-Watts).

An individual FHN unit can be described by the following dimensionless equations~\cite{Fitzhugh-1960a,FitzHugh-1961a,Nagumo-1962a,Cartwright-2000a,Scialla2021a}:
\begin{subequations}\label{eq_FN1}
\begin{align}
%\begin{eqnarray}
\dot{x} &= a \left( x - x^3/3 + y \right) \, ,
	\label{eq_FN1a}
    \\
    \dot{y} &= - \left( x  + by - J \right)/a \, ,
	\label{eq_FN1b}
%\end{eqnarray}
\end{align}
\end{subequations}
where $x$ is the fast activator variable, $y$ is the slow inhibitor variable, and $a$ and $b$ are parameters related to the electrical properties of the unit~\cite{Cartwright-2000a}. 
The value of parameter $J$ determines whether the unit is in the oscillatory regime, which occurs for $|J|< \varepsilon$, or in the excitable one, corresponding to $|J| > \varepsilon$, with $\varepsilon$ defined as
\begin{equation}
\label{epsilon} 
\varepsilon = \frac{3 a^2 - 2 a^2 b -b^2}{3 a^3} \sqrt{a^2 - b} \, . 
\end{equation}

We now build a network of heterogeneous and coupled FHN units, assuming for simplicity that the coupling constant $C$ is the same for all units. 
The FHN equations for the $i$th oscillator in the network then become~\cite{Cartwright-2000a}:
\begin{subequations}\label{eq_FN2}
\begin{align}
    \dot{x}_i &= a \left[x_i - x_i^3/3 + y_i + C \sum_{j} (x_j - x_i)\right] \, ,
	\label{eq_FN2a}
    \\
    \dot{y}_i &= - \left(x_i + b y_i - J_i\right)/a \, ,
	\label{eq_FN2b}
\end{align}
\end{subequations}
where the sum over $j$ in the coupling term in Eq.~\eqref{eq_FN2a}, in the three topologies we will examine, is: 
a) limited to the six nearest neighbors of the $i$th unit, in the case of cubic lattice topology; 
b) running over all network elements, in the case of global coupling; and 
c) limited to the units that are linked to the $i$th element, in the case of Newman-Watts networks. 
Unless otherwise indicated, in all numerical simulation examples presented in this paper, the parameter values in the FHN equations~\eqref{eq_FN2} are set as follows: $a=60$, $b=1.45$, and $C=0.15$. 
The value of the coupling constant $C=0.15$ is the minimum required to observe collective network oscillations and is consistent with results from previous studies~\cite{Scialla2021a}. 
The values of parameters $a$ and $b$ derive from previous work by some of the authors of this paper, aimed at modeling the electrical behavior of $\beta$-cells in the pancreas~\cite{Scialla2021a}. Of course, changing $a$ and $b$ affects not only the value of $\varepsilon$ but also the shape and position of the nullclines of Eqs.~\eqref{eq_FN2}. 
However, the overall qualitative trends described in Sec.~\ref{sec_meanfield} and exemplified in Sec.~\ref{sec_results} do not depend substantially on $a$ and $b$. 

Heterogeneity or diversity is introduced into Eqs.~\eqref{eq_FN2} by associating a different $J_i$ value with each unit $i$. 
In the case of symmetric diversity distributions, we draw $J_i$ values from a Gaussian with mean $\Jav$ and standard deviation $\sigma$. 
Therefore, the latter will determine the degree of diversity of the units constituting the network~\cite{Tessone-2006a,Scialla2021a,Scialla_2022}. 
In order to represent asymmetry and study its effects, as shown in more detail in Sec.~\ref{sec_results}, we  introduce truncations in the normal distribution of $J_i$ values and will also consider half-normal distributions. 
We examine different cases where truncations are introduced both on the sides and in the middle of the above distributions.

%%%%%%%%%%%%%%%%%%%%%%%%%%%%%%%%%%%%%%%%%%%%%%%%%%%%%%%%%%%%%%%%%%%%%%%%5
\section{MEAN-FIELD ANALYSIS}
\label{sec_meanfield}
%%%%%%%%%%%%%%%%%%%%%%%%%%%%%%%%%%%%%%%%%%%%%%%%%%%%%%%%%%%%%%%%%%%%%%%%5

Before presenting numerical results, let us analyze the dynamics of some of the above-mentioned systems through a mean-field approach~\cite{Tessone-2006a,Scialla_2022}, focusing on the differences between symmetric and asymmetric diversity distributions.

Let us introduce the global variables $X(t)=N^{-1}\sum_{i=1}^{N}x_{_{i}}(t)$ and $Y(t)=N^{-1}\sum_{i=1}^{N}y_{_{i}}(t)$. We then use the transformation $x_{_{i}} = X + \delta_{_{i}}$~\cite{desai1978statistical,Tessone-2006a,Scialla_2022,yamakou2022diversity}.
%, assuming that diversity is small 
Upon substitution in the FHN Eqs~\eqref{eq_FN2} we obtain~\cite{Tessone-2006a}
\begin{subequations}
\begin{align}
\dot{X} &= a\Big[X+\delta_i - \frac{(X+\delta_i)^3}{3} + y_i \nonumber \\
       &\qquad + C \sum\left(X + \delta_j - X - \delta_i\right)\Big] \, ,
\label{eq_meanfield1}
\\[1.5ex] % extra vertical space between the two equations
\dot{y}_i &= -\frac{X + \delta_i + b y_i - J_i}{a} \,.
\label{eq_meanfield2}
\end{align}
\end{subequations}
We have not used the transformation $y_{_{i}} = Y + \delta_{_{i}}$, because the $y_{_{i}}$ terms are all linear. 
Also, for simplicity, we only consider the case of all-to-all coupling, therefore the sum in Eq.~\eqref{eq_meanfield1} runs over all $i$'s and $j$'s. 
Developing the cube of the binomial in Eq.~\eqref{eq_meanfield1} we get
\begin{equation}
\begin{split}
\dot{X}=a \Big[X+\delta_i-\frac{X^3+\delta_i^3+3X^2\delta_i+3X\delta_i^2}{3}+y_i  \\
 +C\sum{(X+\delta_j-X-\delta_i) \Big]} \, .
\label{eq_meanfield3}
\end{split}
\end{equation}
Under the assumption that the $J_i$ values follow a zero-mean Gaussian or a symmetric uniform distribution and $\delta_i$'s are small, we can approximate the first and third moments as ${N}^{-1}\sum_{i=1}^{N}{\delta_i\cong0}$ and ${N}^{-1}\sum_{i=1}^{N}{\delta_i^3\cong0}$. 
Therefore, averaging Eq.~\eqref{eq_meanfield3} and Eq.~\eqref{eq_meanfield2} over all ${i}$'s, we obtain
\begin{subequations}
\begin{align}
&\dot{X} = a \left(X-\frac{X^3}{3}-X\frac{1}{N}\sum_{i=1}^{N}\delta_i^2+Y \right) \, ,
\label{global1}
\\
&\dot{Y}=-\frac{X+bY-J_{av}}{a} \, ,
\label{global2}
\end{align}
\end{subequations}
which, using the definition of the second moment, $M\equiv {N}^{-1}\sum_{i=1}^{N}\delta_i^2$~\cite{Tessone-2006a}, can be rewritten as 
\begin{subequations}\label{global}
\begin{align}
&\dot{X}= a \left[X(1-M)-\frac{X^3}{3}+Y \right] \, ,
\label{global4}
\\
&\dot{Y}=-\frac{X+bY-J_{av}}{a} \, .
\label{global5}
\end{align}
\end{subequations}
In Eqs.~\eqref{global}, $M$ expresses the level of diversity, larger $M$ values meaning greater diversity.

\begin{figure}[tb]
	\centering
	\includegraphics[width=0.45\textwidth]{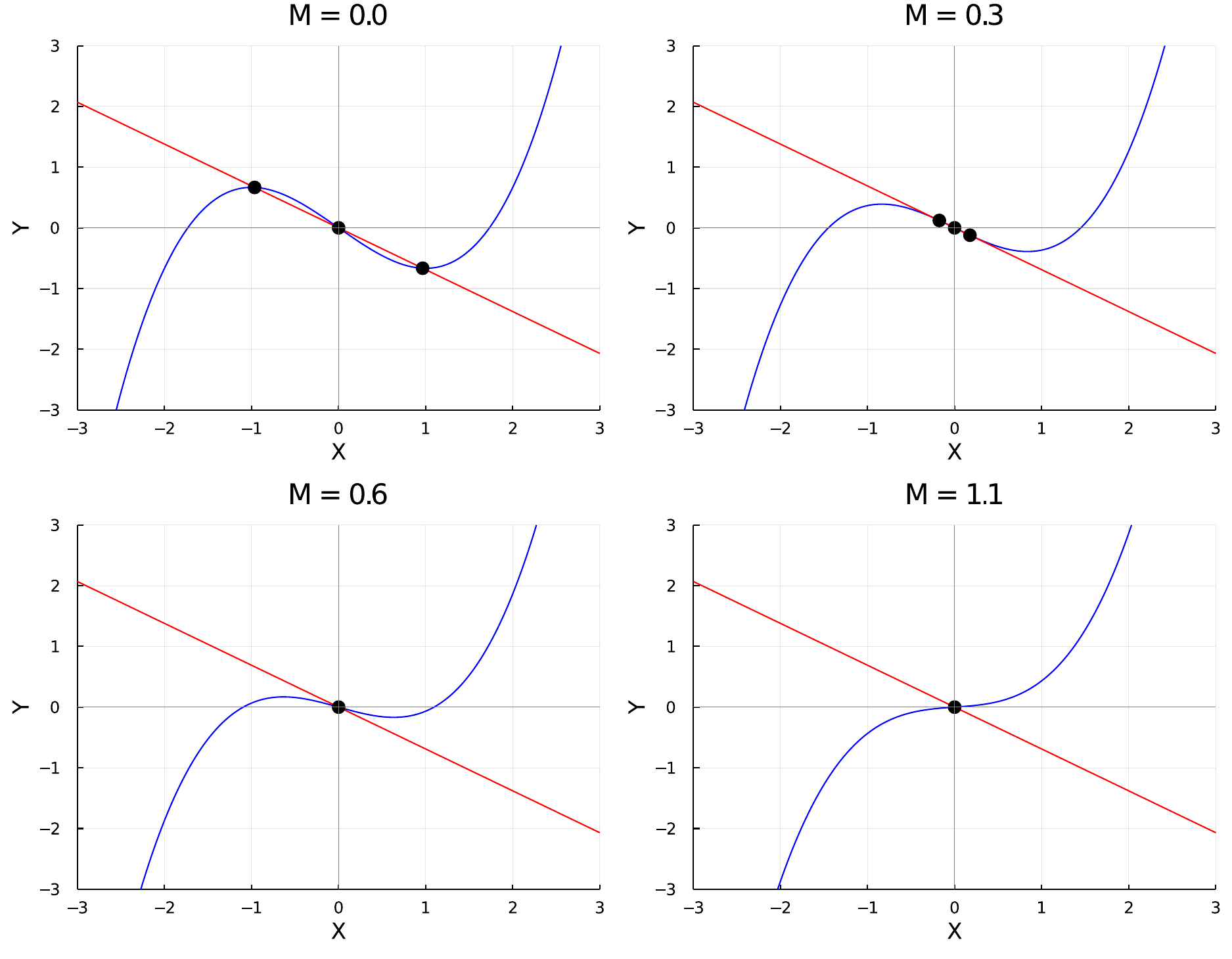}
	\caption{Nullclines of Eqs.~(\ref{global}) for different values of diversity $M$. %Upon increasing $M$. , i.e., diversity, the equilibrium points tend to approach each other on the middle branch of the cubic nullcline, as shown by the comparison between $M=0$ and $M=0.3$, until they merge into a unique equilibrium point, as shown for $M=0.6$. This produces instability and an increased collective oscillatory activity of the network, corresponding to DIR. As $M$ is increased even further, the cubic nullcline becomes monotonic (as shown for $M=1.1$) and the equilibrium point becomes stable, resulting in an excitable, non-oscillatory state of the medium. 
    System parameters are: $a=60$, $b=1.45$, $\Jav=0$.} 
	\label{nullclines}
\end{figure}

In Fig.~\ref{nullclines} we plot the nullclines of Eqs.~(\ref{global}) for different levels of diversity. 
It can be observed that, upon increasing $M$, the equilibrium points of the system tend to approach each other on the middle branch of the cubic nullcline, as shown by the comparison between $M=0$ and $M=0.3$. 
Eventually, as shown for $M=0.6$, the equilibrium points merge, via a subcritical saddle-node bifurcation of fixed points, into a unique unstable equilibrium point, resulting in enhanced global oscillatory activity of the network. 
As $M$ is further increased to $M=1.1$, the middle branch disappears --- the cubic nullcline now becomes monotonic --- so that the unique remaining equilibrium point becomes stable via a subcritical Hopf bifurcation, resulting in an excitable, non-oscillatory medium. 
Therefore, the mean-field analysis predicts that, upon increasing $M$, i.e., diversity, the network should go through an intermediate region of optimal diversity values, characterized by resonant, global oscillatory behavior, followed by a progressive decrease of collective oscillations up to an excitable state in the limit of large $M$. 
This is the DIR effect described in the literature~\cite{Tessone-2006a, Pazo2006f, ShenChuan-Sheng_2008, Toral-2009a,Chen-2009a,Wu-2010a,Wu-2010b,Tessone-2013a,Grace-2014a,Liang-2020a,Scialla2021a, Scialla_2022}.

However, as pointed out above, the approximations $\sum_{i=1}^{N}{\delta_i\cong0}$ and $\sum_{i=1}^{N}{\delta_i^3\cong0}$ are valid only under the assumption that the distributions of $J_i$ values are Gaussian or uniform, and symmetric about a zero mean. 
If this condition is not met, the corresponding terms in Eq.~\eqref{eq_meanfield3} cannot be neglected. 
The presence of these terms determines a shift of the equilibrium point(s) toward the ascending or descending branch of the cubic nullcline, depending on their sign, with the result that a diversity increase now tends to make the network excitable rather than oscillatory. 
In addition, if asymmetry causes a significant shift of the mean value $\Jav$ of the $J_i$ distribution away from its original position, that can also contribute to moving the equilibrium points toward their stability region. 

The above considerations, although qualitative, illustrate that the degree of symmetry of the diversity distribution has a crucial impact on the network's ability to exhibit global oscillations. 
This effect can explain the behavior of a wide range of network configurations, including those where the ratio between the number of oscillatory and excitable units is not an applicable criterion.

%%%%%%%%%%%%%%%%%%%%%%%%%%%%%%%%%%%%%%%%%%%%%%%%%%%%%%%%%%%%%%%%%%%%%%%%5
\section{RESULTS}
\label{sec_results}
%%%%%%%%%%%%%%%%%%%%%%%%%%%%%%%%%%%%%%%%%%%%%%%%%%%%%%%%%%%%%%%%%%%%%%%%5

We now examine a series of network examples with varying diversity distributions and topologies to illustrate how the diversity distribution symmetry affects the global oscillatory activity of the network. Although numerical results are shown for only a subset of topologies, we consistently observed the same outcome, in terms of presence or absence of global oscillations, for each diversity distribution across all three tested configurations: all-to-all coupling, cubic lattice, and cubic lattice augmented with small-world features. The latter was implemented according to the Newman–Watts model, with rewiring probabilities ranging from 0 to 1 \cite{NEWMAN1999341}.

%%%%%%%%%%%%%%%%%%%%%%%%%%%%%%%%%%%%%%%%%%%%%%%%%%%%%%%%%%%%%%%%%%%%%%%%5
\subsection{Symmetry metrics}
\label{sec_METR}
%%%%%%%%%%%%%%%%%%%%%%%%%%%%%%%%%%%%%%%%%%%%%%%%%%%%%%%%%%%%%%%%%%%%%%%%5

Let us first introduce some metrics to quantify the degree of symmetry of the various bias distributions we are considering. We will utilize two parameters: a ``normalized center of mass'' (nCOM) and a ``symmetry balance score'' (SBS).

We define the nCOM of a distribution as
\begin{equation}
\text{nCOM} \equiv \frac{\left| \sum_{i=1}^N (J_i - J_0) \right|}{N \cdot \varepsilon} \,,
\end{equation}
where $N$ is the number of elements in the distribution; $J_{0}$ is the center of the oscillatory interval, which in our examples is $J_{0}=0$ since the oscillatory interval is $(-\varepsilon,+\varepsilon)$; and $\varepsilon$, defined in Eq.~(\ref{epsilon}), provides a normalization factor that makes nCOM independent of the width of the oscillatory interval. 
The nCOM quantifies the aggregate shift of the distribution with respect to the center of the oscillatory range and takes values in the interval $[0,\infty)$. 
Values of $\text{nCOM} \in [0,1)$ correspond to a fairly symmetric (or moderately asymmetric) distribution centered in the oscillatory range $(-\varepsilon,+\varepsilon)$, whereas $\text{nCOM} \in [1,\infty)$ indicates increasing asymmetry, with the mass of the distribution predominantly concentrated on one side of the center and outside the oscillatory range. 
Therefore, we expect that distributions with $\text{nCOM} \in [0,1)$ should correspond to networks characterized by sustained collective oscillations, whereas $\text{nCOM} \geq1$ should be indicative of the absence of global oscillatory activity.

The SBS is instead defined as
\begin{equation}
\text{SBS}\equiv \frac{\min(N_+, N_-)}{\max(N_+, N_-)} \,,
\end{equation}
where $N_+$ and $N_-$ denote, respectively, the number of elements in the distribution such that $J_i > J_{0}$ and $J_i < J_{0}$. 
The SBS, which can be viewed as an adaptation of the imbalance ratio used in classification problems~\cite{He2009}, captures the balance in the number of components on either side of $J_{0}$ and can vary between 0 and 1. 
A value of $\text{SBS} = 1$ indicates a perfectly symmetric distribution with equal populations on both sides of $J_{0}$, while values equal to or close to zero correspond to highly imbalanced (asymmetric) distributions. 
Consequently, we expect bias distributions with very low or zero SBS to be unable to exhibit global network oscillations, while SBS values significantly higher than zero and up to 1 should give rise to collective oscillations.

Together, these two metrics provide complementary information: nCOM reflects the net displacement of the distribution's mass from the center of the oscillatory range, while SBS evaluates whether the distribution is balanced in terms of the number of components on each side. 
However, it should be noted that while nCOM can be calculated for any distribution, SBS is only applicable to distributions that have at least one element on either side of $J_{0}$. For instance, if $J_{0}=0$ and all $J_i$'s are positive, SBS would always assume the same value (SBS=0) regardless of the shape of the distribution, therefore it is not meaningful in such cases.

It should be noted that although all FHN networks considered in this study have an oscillatory interval centered at zero ($J_0 = 0$), our conclusions regarding the effects of distribution symmetry remain valid for any value of $J_0$ that may arise in different formulations of the FHN equations.

\begin{table*}[t]
\centering
\caption{Comparison of symmetry metrics and oscillatory behavior for the nine bias distributions corresponding to Figs~2--10. Green indicates correct predictions, while red highlights discrepancies. The criterion for global network oscillations is shown in parentheses in each column heading.}
\scriptsize
\renewcommand{\arraystretch}{1.7}
\setlength{\tabcolsep}{4pt}
\begin{tabular}{ >{\centering\arraybackslash}m{1.2cm}
                 >{\centering\arraybackslash}m{3.1cm}
                 >{\centering\arraybackslash}m{2.8cm}
                 >{\centering\arraybackslash}m{2.6cm}
                 >{\centering\arraybackslash}m{3.2cm}
                 >{\centering\arraybackslash}m{1.8cm} }
\hline
\textbf{Figure} &
\textbf{Distribution} &
\shortstack{\textbf{Oscillator/}\\\textbf{Nonoscillator Ratio}\\\textbf{($> 0.3$)}} &
\shortstack{\textbf{Normalized}\\\textbf{Center of Mass}\\\textbf{($< 1$)}} &
\shortstack{\textbf{Symmetry Balance}\\\textbf{Score ($\gg 0$)}} &
\textbf{Global Oscillations?} \\
\hline
2 &
\raisebox{-.5\height}{\includegraphics[width=3.0cm]{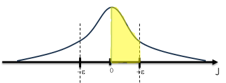}}\rule{0pt}{2.4em} &
\textcolor{green!80!black}{\textbf{Inf.}} &
\textcolor{green!80!black}{\textbf{0.50}} &
n.a. &
Yes \\
3 &
\raisebox{-.5\height}{\includegraphics[width=3.0cm]{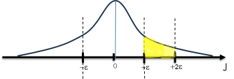}}\rule{0pt}{2.4em} &
\textcolor{green!80!black}{\textbf{0}} &
\textcolor{green!80!black}{\textbf{1.50}} &
n.a. &
No \\
4 &
\raisebox{-.5\height}{\includegraphics[width=3.0cm]{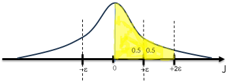}}\rule{0pt}{2.4em} &
\textcolor{green!80!black}{\textbf{1}} &
\textcolor{green!80!black}{\textbf{0.99}} &
n.a. &
Yes \\
5 &
\raisebox{-.5\height}{\includegraphics[width=3.0cm]{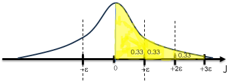}}\rule{0pt}{2.4em} &
\textcolor{red}{\textbf{0.5}} &
\textcolor{green!80!black}{\textbf{1.52}} &
n.a. &
No \\
6 &
\raisebox{-.5\height}{\includegraphics[width=3.0cm]{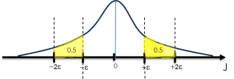}}\rule{0pt}{2.4em} &
\textcolor{red}{\textbf{0}} &
\textcolor{green!80!black}{\textbf{0.00}} &
\textcolor{green!80!black}{\textbf{0.95}} &
Yes \\
7 &
\raisebox{-.5\height}{\includegraphics[width=3.0cm]{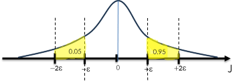}}\rule{0pt}{2.4em} &
\textcolor{red}{\textbf{0}} &
\textcolor{red}{\textbf{1.37}} &
\textcolor{red}{\textbf{0.045}} &
Yes \\
8 &
\raisebox{-.5\height}{\includegraphics[width=3.0cm]{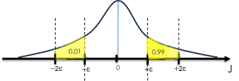}}\rule{0pt}{2.4em} &
\textcolor{green!80!black}{\textbf{0}} &
\textcolor{green!80!black}{\textbf{1.47}} &
\textcolor{green!80!black}{\textbf{0.015}} &
No \\
9 &
\raisebox{-.5\height}{\includegraphics[width=3.0cm]{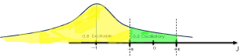}}\rule{0pt}{2.4em} &
\textcolor{green!80!black}{\textbf{0.25}} &
\textcolor{green!80!black}{\textbf{25.15}} &
\textcolor{green!80!black}{\textbf{0.00}} &
No \\
10 &
\raisebox{-.5\height}{\includegraphics[width=3.0cm]{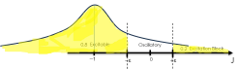}}\rule{0pt}{2.4em} &
\textcolor{red}{\textbf{0}} &
\textcolor{red}{\textbf{24.01}} &
\textcolor{green!80!black}{\textbf{0.26}} &
Yes \\
\hline
\end{tabular}
\label{tab:asymmetry_summary}
\end{table*}

Table~\ref{tab:asymmetry_summary} reports the nCOM and SBS values for all the diversity distributions that will be considered in the following sections, alongside the corresponding ratios of oscillatory to non-oscillatory units, that is, the main criterion considered in previous literature. 
Beyond being strictly applicable only to a narrow subset of distributions, a key limitation of this criterion is the absence of a clear threshold ratio, above which global network oscillations are expected to emerge. 
In Table~\ref{tab:asymmetry_summary}, we have assumed this threshold to be $\approx 0.3$, allowing a correct prediction for the distribution shown in Fig.~\ref{final1}. 
However, there are examples from the previous literature where an oscillator/non-oscillator ratio of 0.25 resulted in collective network oscillations~\cite{Pazo2006f}.

%Looking at the symmetry parameter values reported in Table~\ref{tab:asymmetry_summary} for the five distributions examined in this Subsec.~\ref{sec_DIR}, we can observe that the distributions in Fig.~\ref{halfnormal1} and Fig.~\ref{halfnormal3} have nCOM values lower than 1 and, as such, give rise to global network oscillations. 
%Instead, those in Fig.~\ref{halfnormal2} and Fig.~\ref{halfnormal4}, which are less symmetric, have $\text{nCOM}>1$ and, accordingly, do not produce sustained collective oscillations. 
%Therefore, nCOM is able to correctly predict all four cases, whereas the oscillator fraction criterion fails to explain the lack of global oscillations for the distribution corresponding to Fig.~\ref{halfnormal4}.

%%%%%%%%%%%%%%%%%%%%%%%%%%%%%%%%%%%%%%%%%%%%%%%%%%%%%%%%%%%%%%%%%%%%%%%%5
\subsection{Half-normal diversity distributions}
\label{sec_DIR}
%%%%%%%%%%%%%%%%%%%%%%%%%%%%%%%%%%%%%%%%%%%%%%%%%%%%%%%%%%%%%%%%%%%%%%%%5

\begin{figure}[tb]
	\centering
	\includegraphics[width=0.45\textwidth]{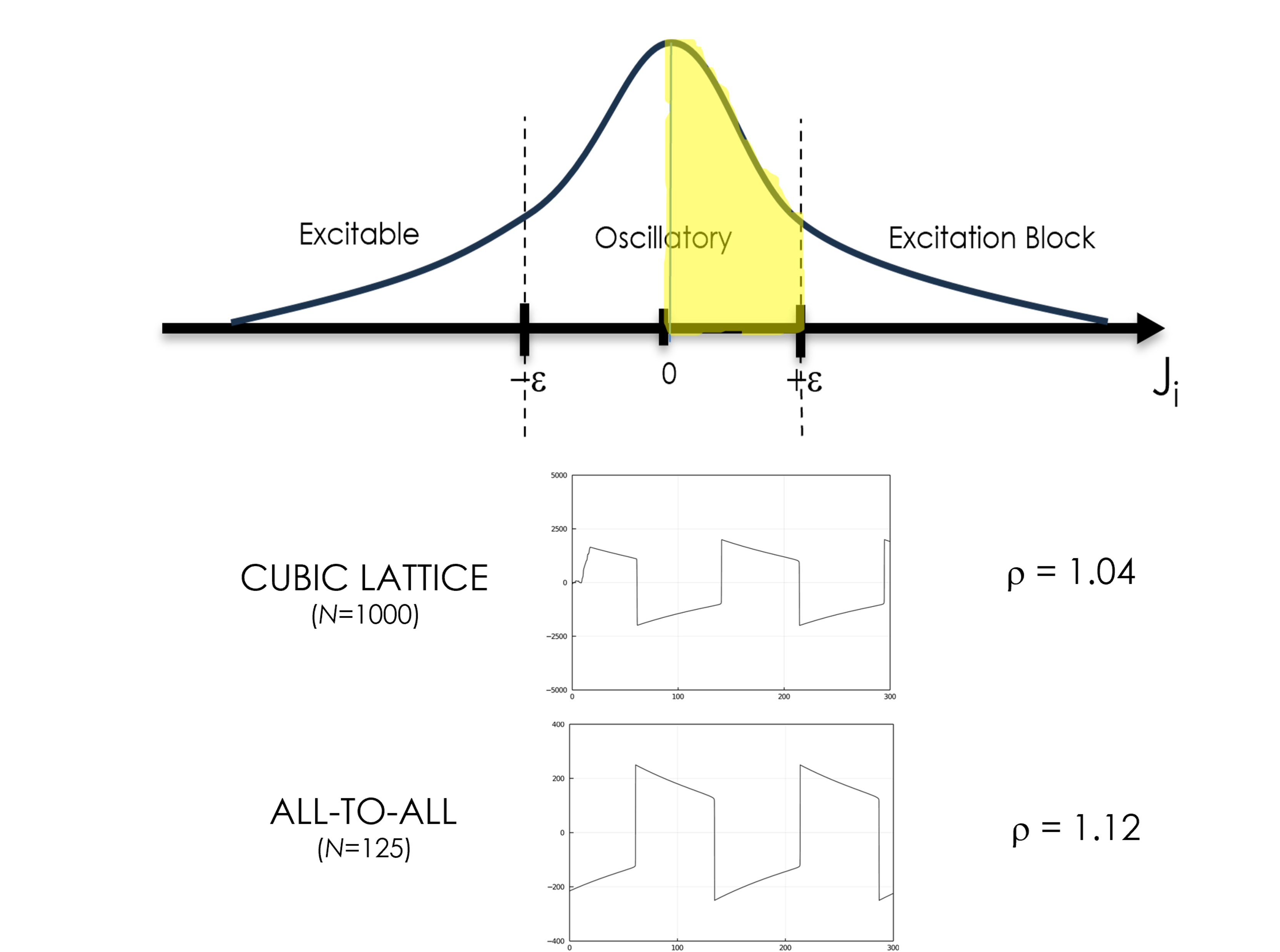}
	\caption{Truncated half-normal diversity distribution comprising oscillatory units only ($\Jav=0$, $\sigma=0.5$). The yellow area highlights the portion of the Gaussian that has been used to sample $J_i$ values. In this case, the lack of symmetry has no negative impact on global network oscillations. The nCOM parameter (nCOM = 0.50,  Table~\ref{tab:asymmetry_summary}) correctly predicts collective oscillations.} 
	\label{halfnormal1}
\end{figure}

We begin our analysis by studying half-normal diversity distributions and take as a first example a truncated distribution, with $\Jav=0$ and $\sigma=0.5$, where $J_i$ values are drawn exclusively from the interval $[0, \varepsilon)$. 
This means that all network units are individually in an oscillatory state and the diversity distribution is relatively asymmetric, because we are picking $J_i$ values from the positive semiaxis only (Fig.~\ref{halfnormal1}).

After numerically solving the FHN equations \eqref{eq_FN2} for the cubic lattice and all-to-all topology, we compute the global oscillatory activity of the network, $\rho$, from the expression~\cite{Cartwright-2000a,Scialla_2022}
\begin{equation}
\rho\equiv N^{-2} \sqrt{\left \langle [S(t) -\bar{S}]^2 \right \rangle} \, ,
\label{FNENG7}
\end{equation}
where $N$ is the total number of units, $S(t) = \sum_i x_i(t)$, and $\bar{S} = \langle S(t) \rangle$, with $\langle \dots \rangle$ denoting a time average. A strong collective oscillatory activity corresponds to $\rho \approx 1$, while $\rho \approx 0$ indicates no activity.

As shown in Fig.~\ref{halfnormal1}, a network with the above diversity distribution exhibits strong collective oscillations, in the cases of both cubic lattice and all-to-all topology. Therefore, asymmetry does not prevent a network made of oscillatory units only from being in a globally oscillatory state.

\begin{figure}[tb]
	\centering
	\includegraphics[width=0.45\textwidth]{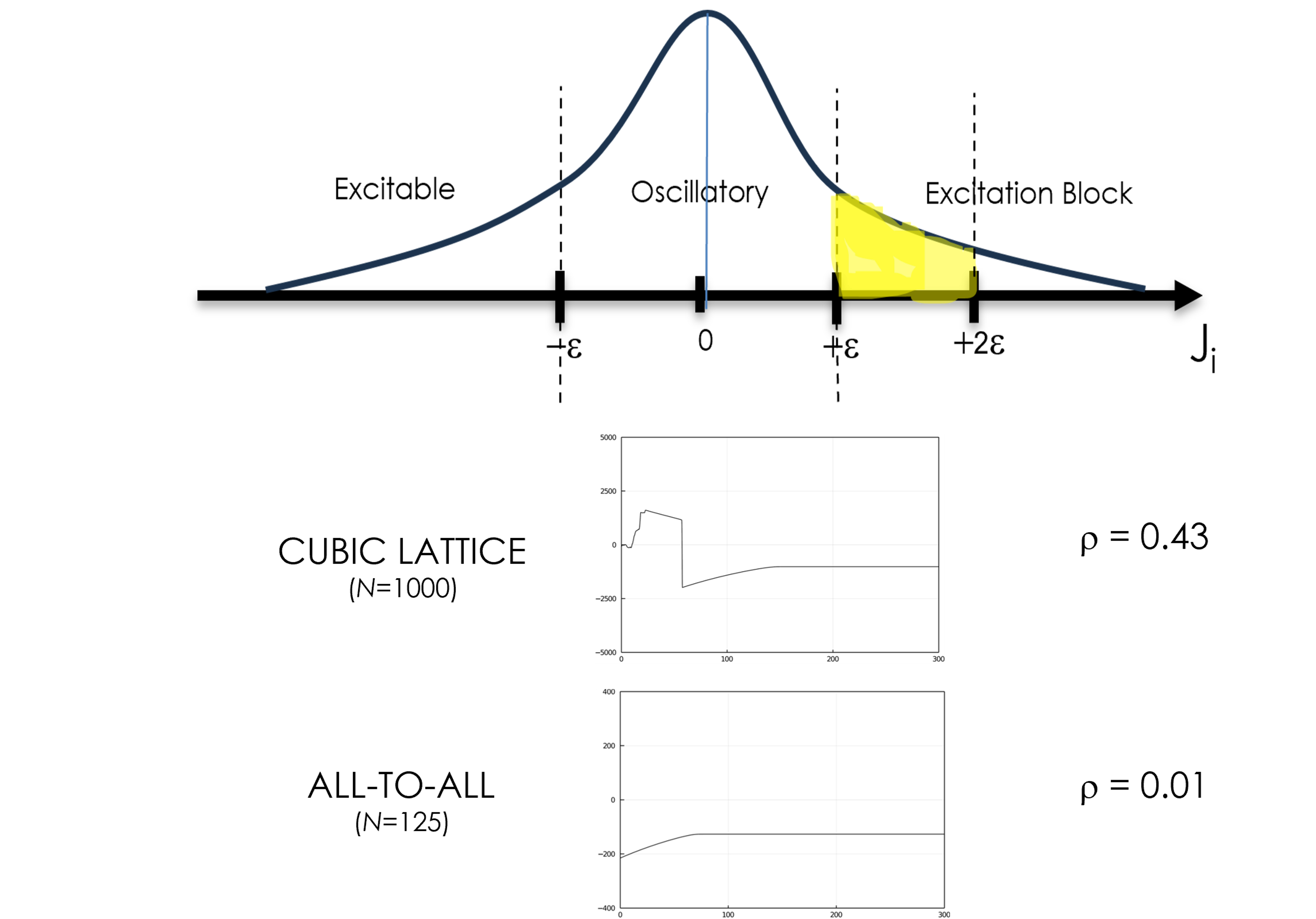}
	\caption{Truncated half-normal diversity distribution comprising excitable units only. The yellow area highlights the portion of the Gaussian that has been used to sample $J_i$ values. In this case, the lack of symmetry prevents global network oscillations. The nCOM parameter (nCOM = 1.50, Table~\ref{tab:asymmetry_summary}) correctly predicts the absence of global oscillations.} 
	\label{halfnormal2}
\end{figure}

Let us now examine the behavior of a network whose constituent units are still picked exclusively from the positive semiaxis, but all of them are individually in an excitable state, since they belong to the interval $(\varepsilon, 2\varepsilon]$  (Fig.~\ref{halfnormal2}).
In this case, as shown in Fig.~\ref{halfnormal2}, the asymmetry in the diversity distribution is able to prevent collective network oscillations for both studied topologies.

\begin{figure}[tb]
	\centering
	\includegraphics[width=0.45\textwidth]{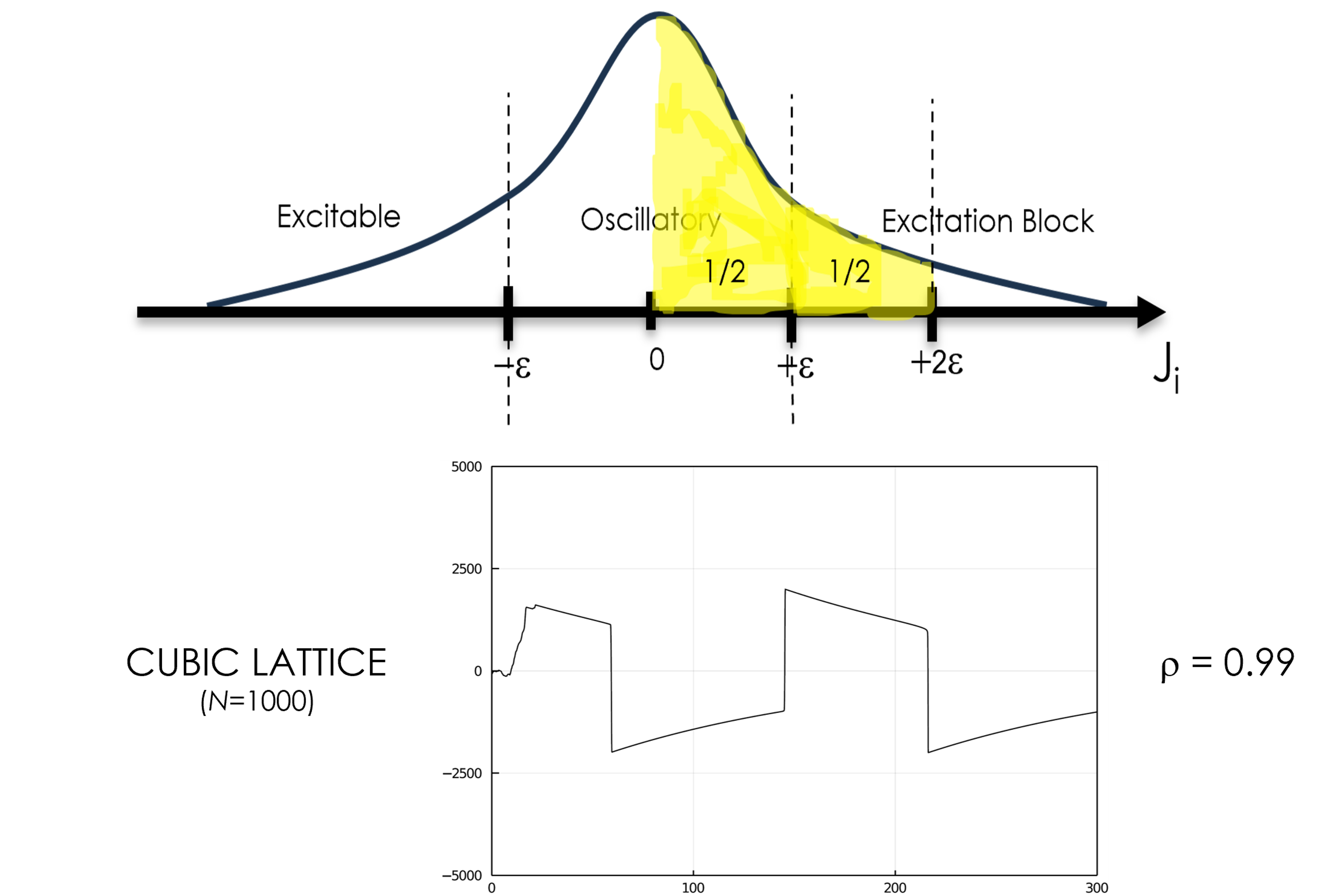}
	\caption{Truncated half-normal diversity distribution comprising both oscillatory and excitable units in a $50/50\%$ ratio. The yellow area highlights the portion of the Gaussian that has been used to sample $J_i$ values. In this case, the lack of symmetry has no negative impact on global network oscillations. The nCOM parameter (nCOM = 0.99, Table~\ref{tab:asymmetry_summary}) correctly predicts this.} 
	\label{halfnormal3}
\end{figure}

\begin{figure}[tb]
	\centering
	\includegraphics[width=0.45\textwidth]{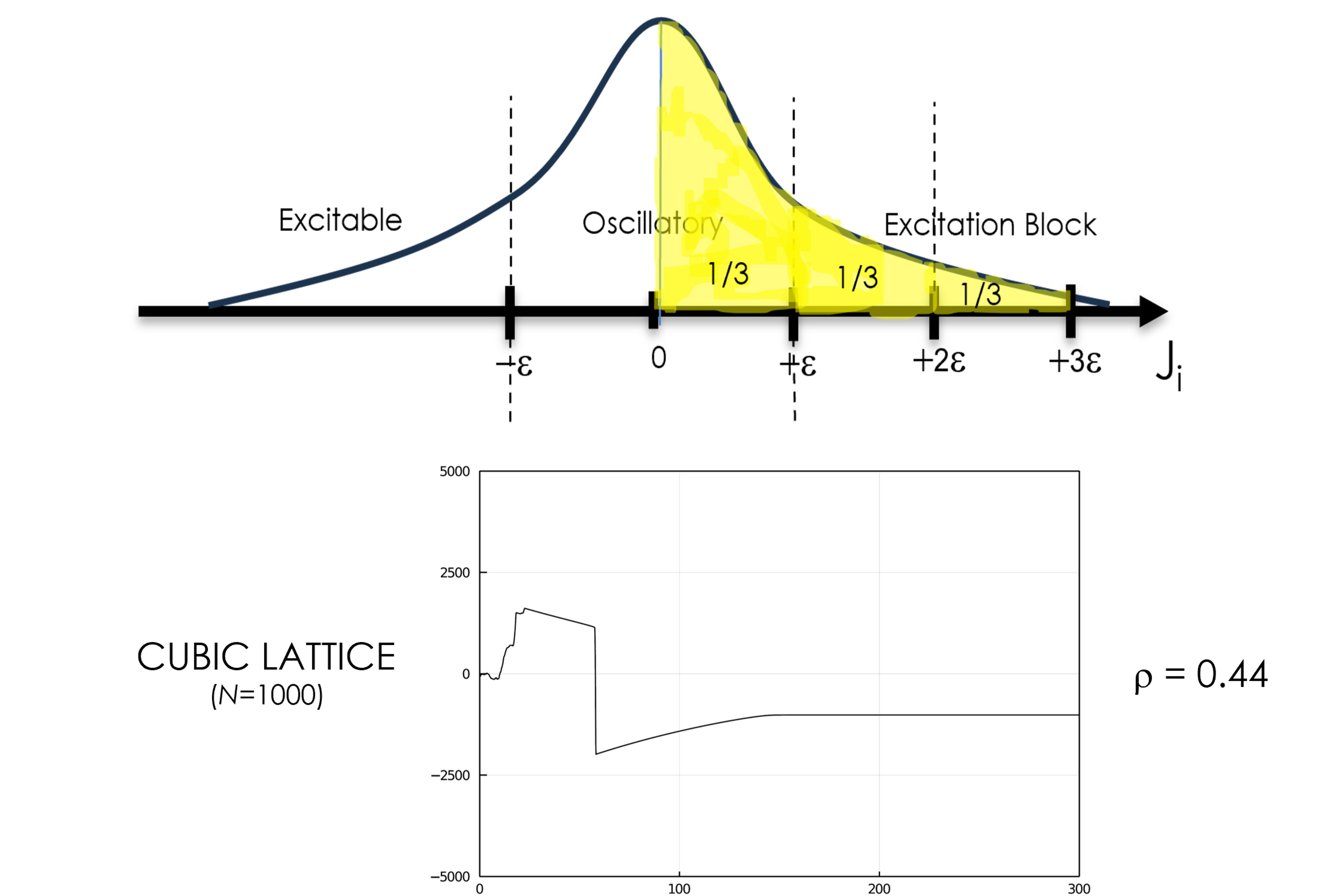}
	\caption{Truncated half-normal diversity distribution comprising both oscillatory and excitable units in a 1/2  oscillatory/excitable ratio. The yellow area highlights the portion of the Gaussian that has been used to sample $J_i$ values. In this case, the degree of distribution asymmetry is such that it determines the absence of sustained global network oscillations. The nCOM parameter (nCOM = 1.52, Table~\ref{tab:asymmetry_summary}) correctly predicts this.} 
	\label{halfnormal4}
\end{figure}

But what happens in the case of a truncated, half-normal diversity distribution that includes both excitable and oscillatory units? For instance, let us consider the diversity distribution shown in Fig.~\ref{halfnormal3}. Here, half of the units are picked from the interval $[0,\varepsilon)$ and are therefore individually oscillatory, while the remaining half belong to the interval $(\varepsilon, 2\varepsilon]$ and as such are excitable.
As shown in Fig.~\ref{halfnormal3}, the outcome in terms of global network oscillations is similar to that in Fig.~\ref{halfnormal1}, i.e., the network is in a resonant oscillatory state even though the diversity distribution is asymmetric and there is a substantial number of units that are individually non-oscillatory. Here we show the result for the cubic lattice topology only; however, the all-to-all coupling case exhibits the same behavior. 
It is worth pointing out that this outcome is consistent with previous studies reported in the literature~\cite{Tessone-2006a,Pazo2006f,ShenChuan-Sheng_2008}, which explained this dynamics on the basis of a mechanism where the oscillatory units pull the other units, generating global oscillations. In line with this reasoning, if we now add more excitable units to the diversity distribution, achieving a 2/1 ratio of excitable to oscillatory units, as shown in Fig.~\ref{halfnormal4}, we observe that global network oscillations disappear. According to the pull mechanism, this happens because the relative amount of non-oscillatory units has now become too high. However, we should also point out that the diversity distribution shown in Fig.~\ref{halfnormal4} is more asymmetric than the one in Fig.~\ref{halfnormal3}.

Looking at the symmetry parameter values reported in Table~\ref{tab:asymmetry_summary} for the five distributions examined in this subsection, we can observe that the distributions in Fig.~\ref{halfnormal1} and Fig.~\ref{halfnormal3} have nCOM values lower than 1 and, as such, give rise to global network oscillations. 
Instead, those in Fig.~\ref{halfnormal2} and Fig.~\ref{halfnormal4}, which are less symmetric, have $\text{nCOM}>1$ and, accordingly, do not produce sustained collective oscillations. 
Therefore, nCOM is able to correctly predict all four cases, whereas the oscillator fraction criterion fails to explain the lack of global oscillations for the distribution corresponding to Fig.~\ref{halfnormal4}.

%%%%%%%%%%%%%%%%%%%%%%%%%%%%%%%%%%%%%%%%%%%%%%%%%%%%%%%%%%%%%%%%%%%%%%%%%%%%%%
\subsection{Normal diversity distributions}
\label{sec_HUB}
%%%%%%%%%%%%%%%%%%%%%%%%%%%%%%%%%%%%%%%%%%%%%%%%%%%%%%%%%%%%%%%%%%%%%%%%5

We now turn to examples of truncated normal distributions of $J_i$ values, i.e., diversity distributions characterized by the presence of $J_i$ values on both sides of the mode of the parent, non-truncated Gaussian.

\begin{figure}[tb]
	\centering
	\includegraphics[width=0.45\textwidth]{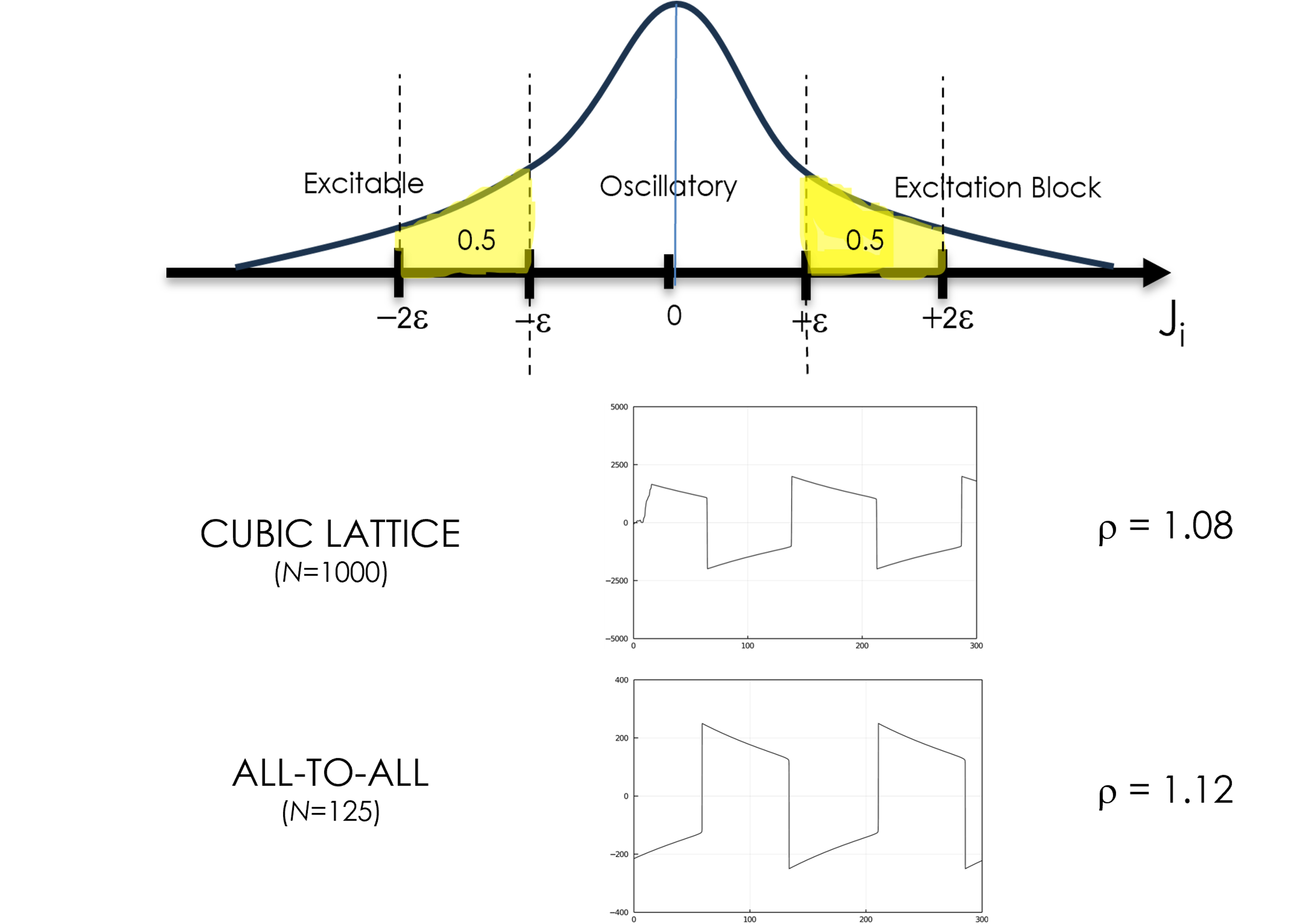}
	\caption{Truncated normal diversity distribution comprising excitable units only. The units are symmetrically distributed on the two sides of the mode in a 1/1 ratio. The yellow area highlights the portion of the Gaussian that has been used to sample $J_i$ values. The distribution is fully symmetric and gives rise to global network oscillations in both studied topologies. The values of nCOM and SBS (nCOM = 0, SBS = 0.95, Table~\ref{tab:asymmetry_summary}) are fully consistent with this result, while the criterion based on counting the fraction of oscillatory units predicts no collective oscillations.} 
	\label{normal1}
\end{figure}

The first case we analyze is an extension of the system presented in Ref.~\cite{Cartwright-2000a}. 
In that study, the network population was divided into two subgroups, made of identical units with $J_i$ values below and above the oscillatory interval $-\varepsilon<J_i<\varepsilon$, respectively. 
Here, we consider a diversity distribution made of excitable units only, which are symmetrically positioned on the two sides of the mode of a Gaussian, as shown in Fig.~\ref{normal1}. 
This distribution is fully symmetric, with $\text{nCOM}=0$ and $\text{SBS} = 0.95$ (Table~\ref{tab:asymmetry_summary}), and produces strong collective network oscillations, just like the system in Ref.~\cite{Cartwright-2000a}. 
There is no way to explain or predict the oscillatory behavior of this network based on the ratio between oscillatory and excitable units, whereas the observed dynamics is fully consistent with our considerations about the symmetry of the diversity distribution.

\begin{figure}[tb]
	\centering
	\includegraphics[width=0.45\textwidth]{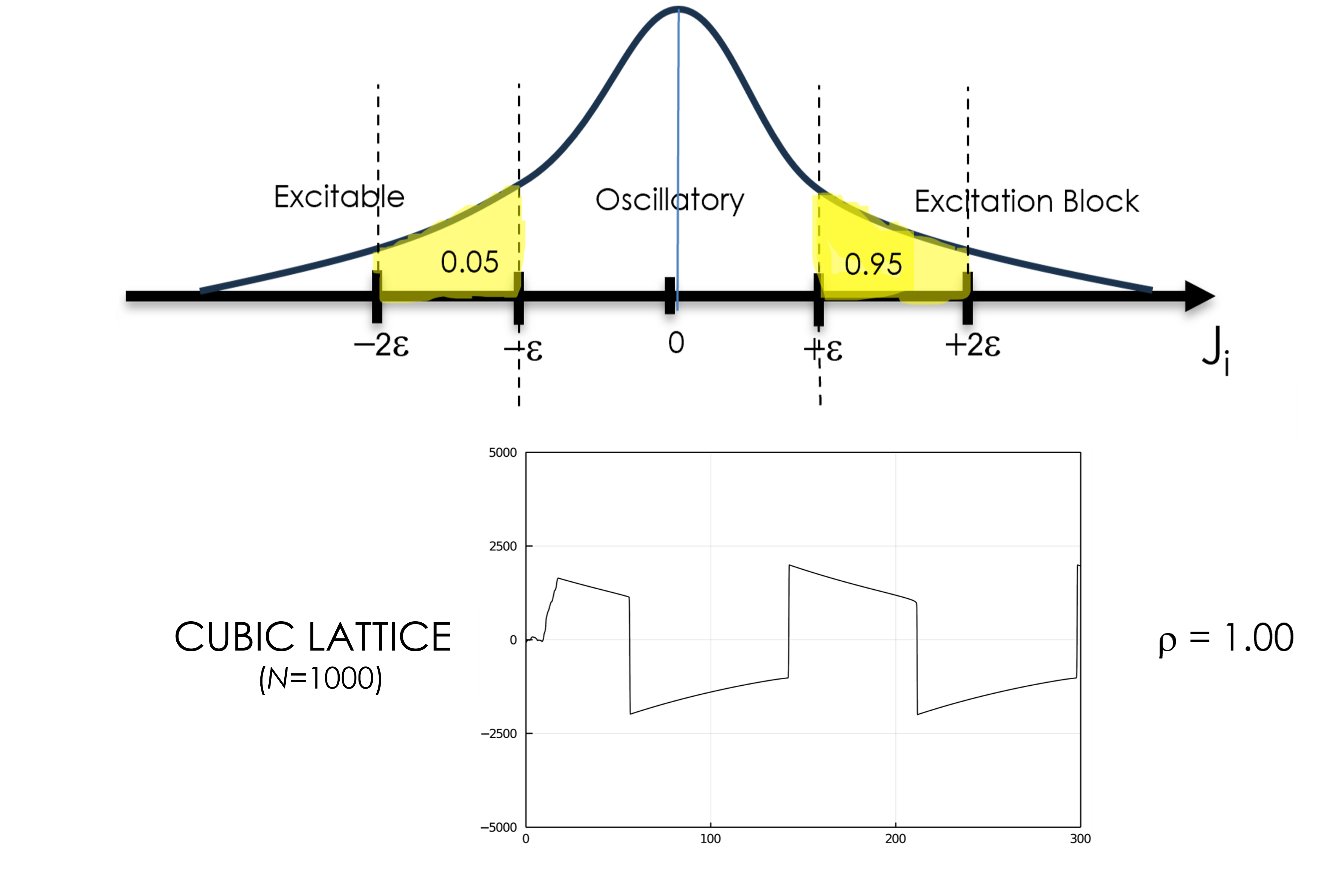}
	\caption{Truncated normal diversity distribution comprising excitable units only. The units are distributed on the two sides of the mode in a 5/95\% ratio. The yellow area highlights the portion of the Gaussian that has been used to sample $J_i$ values. The distribution has a certain degree of asymmetry yet it gives rise to global network oscillations. Both nCOM and the oscillator fraction criterion fail to predict this result, however, nCOM correctly ranks this distribution as more likely to produce global oscillations with respect to Fig.~\ref{halfnormal2}, Fig.~\ref{halfnormal4}, and Fig.~\ref{normal3}. SBS correctly shows that this distribution is more likely to generate collective oscillations than that in Fig.~\ref{normal3} (Table~\ref{tab:asymmetry_summary}).}
	\label{normal2}
\end{figure}

\begin{figure}[tb]
	\centering
	\includegraphics[width=0.45\textwidth]{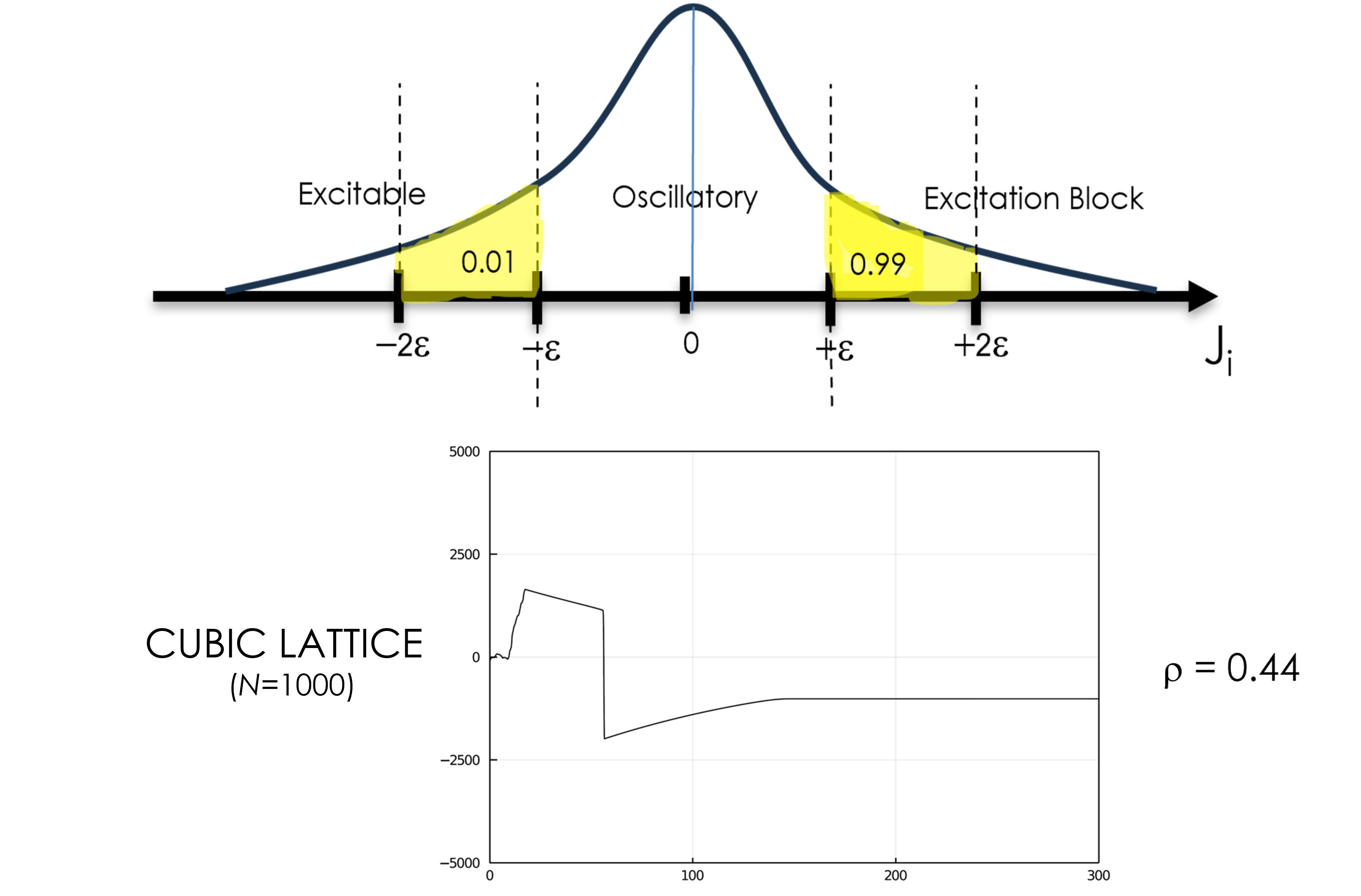}
	\caption{Truncated normal diversity distribution comprising excitable units only. The units are distributed on the two sides of the mode in a 1/99\% ratio. The yellow area highlights the portion of the Gaussian that has been used to sample $J_i$ values. The degree of asymmetry of this distribution is such that the network has no global oscillations, as predicted by all parameters in Table~\ref{tab:asymmetry_summary}.} 
	\label{normal3}
\end{figure}

We now introduce a degree of asymmetry into the diversity distribution of Fig.~\ref{normal1}, by picking a different number of units from either side of the truncated Gaussian. Specifically, a fraction of the total number of units equal to $0.95$ are taken from the positive side and 0.05 from the negative one (Fig.~\ref{normal2}). 
In Fig.~\ref{normal3} we push this asymmetry even further, by picking 0.99 of the units from the positive side and only $0.01$ from the negative one.


Simulation results show that even with a 95/5\% ratio of positive to negative $J_i$ values, the network is still capable of strong global oscillations, whereas at a 99/1\% ratio this does not happen any more. 
Upon comparing the diversity distributions in Fig.~\ref{normal2} %($\text{nCOM}=1.37$, $\text{SBS} = 0.045$) 
and Fig.~\ref{normal3} %($\text{nCOM}=1.47$, $\text{SBS} = 0.015$) 
to that in Fig.~\ref{halfnormal2}, %($\text{nCOM}=1.50$, $\text{SBS} = n.a.$), 
we should point out that the latter is more asymmetric than both of the former, as shown by its higher nCOM value (Table~\ref{tab:asymmetry_summary}). 
Among these three distributions, only the one with the lowest nCOM value, i.e., with the highest degree of symmetry (corresponding to Fig.~\ref{normal2}), is able to produce sustained collective oscillations. 
Notice that although the nCOM parameter makes the incorrect prediction that the distribution in Fig.~\ref{normal2} should not exhibit global oscillations because the criterion $\text{nCOM}<1$ is not satisfied, it still correctly indicates that the relative tendency to global oscillations should be in the order $\text{Fig.}~\ref{normal2} > \text{Fig.}~\ref{normal3} > \text{Fig.}~\ref{halfnormal2}$, in line with the increasing symmetry (i.e., decreasing nCOM values) of the distributions.

\begin{figure}[tb]
	\centering
	\includegraphics[width=0.45\textwidth]{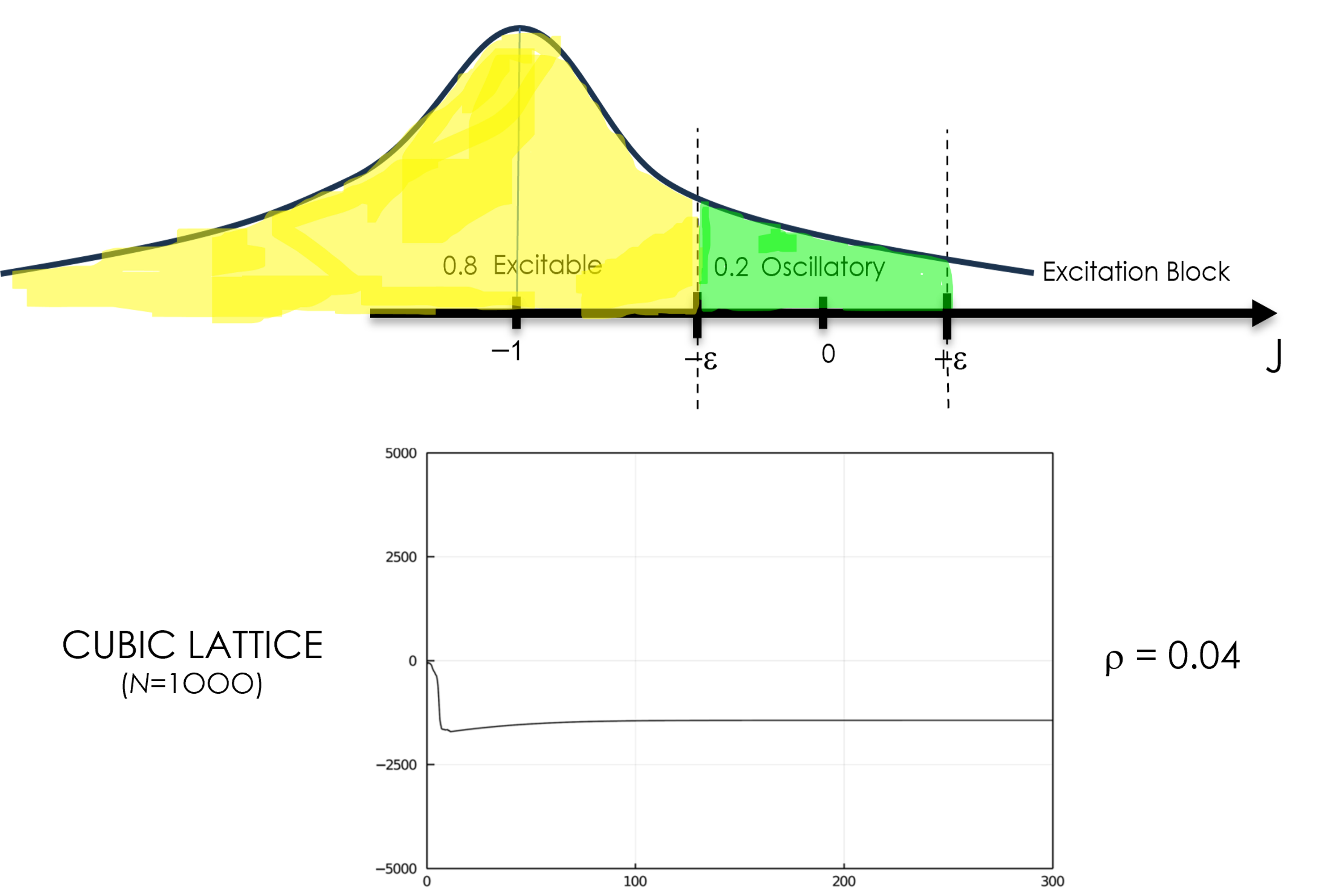}
	\caption{Truncated normal diversity distribution comprising both excitable and oscillatory units. Excitable units have $J_i$ values distributed over the range $(-\infty,-\varepsilon)$, while $J_i$ values for oscillatory units are picked from $(-\varepsilon,+\varepsilon)$. The yellow area highlights the portion of the Gaussian that has been used to sample $J_i$ values. The asymmetry of this distribution results in the absence of collective oscillations, as predicted by all parameters in Table~\ref{tab:asymmetry_summary}.} 
	\label{final1}
\end{figure}

As a final example of normal diversity distribution, we study a case where the mode of the distribution is subthreshold, i.e., below the interval of oscillatory $J_i$ values, Fig.~\ref{final1} (in all previous cases, the mode of the distribution was positioned at the center of the oscillatory interval). 
Excitable units have $J_i$ values distributed over the range $(-\infty, -\varepsilon)$, while $J_i$ values for oscillatory units are picked from $(-\varepsilon,+\varepsilon)$. 
Here we see again that the network has no collective oscillatory activity, which can be explained by both the criterion based on the numerical ratio between oscillatory and excitable units (too low in this case), and considerations related to the degree of symmetry of the diversity distribution, which is also very low ($\text{nCOM} = 25.15$).

\begin{figure}[tb]
	\centering
	\includegraphics[width=0.45\textwidth]{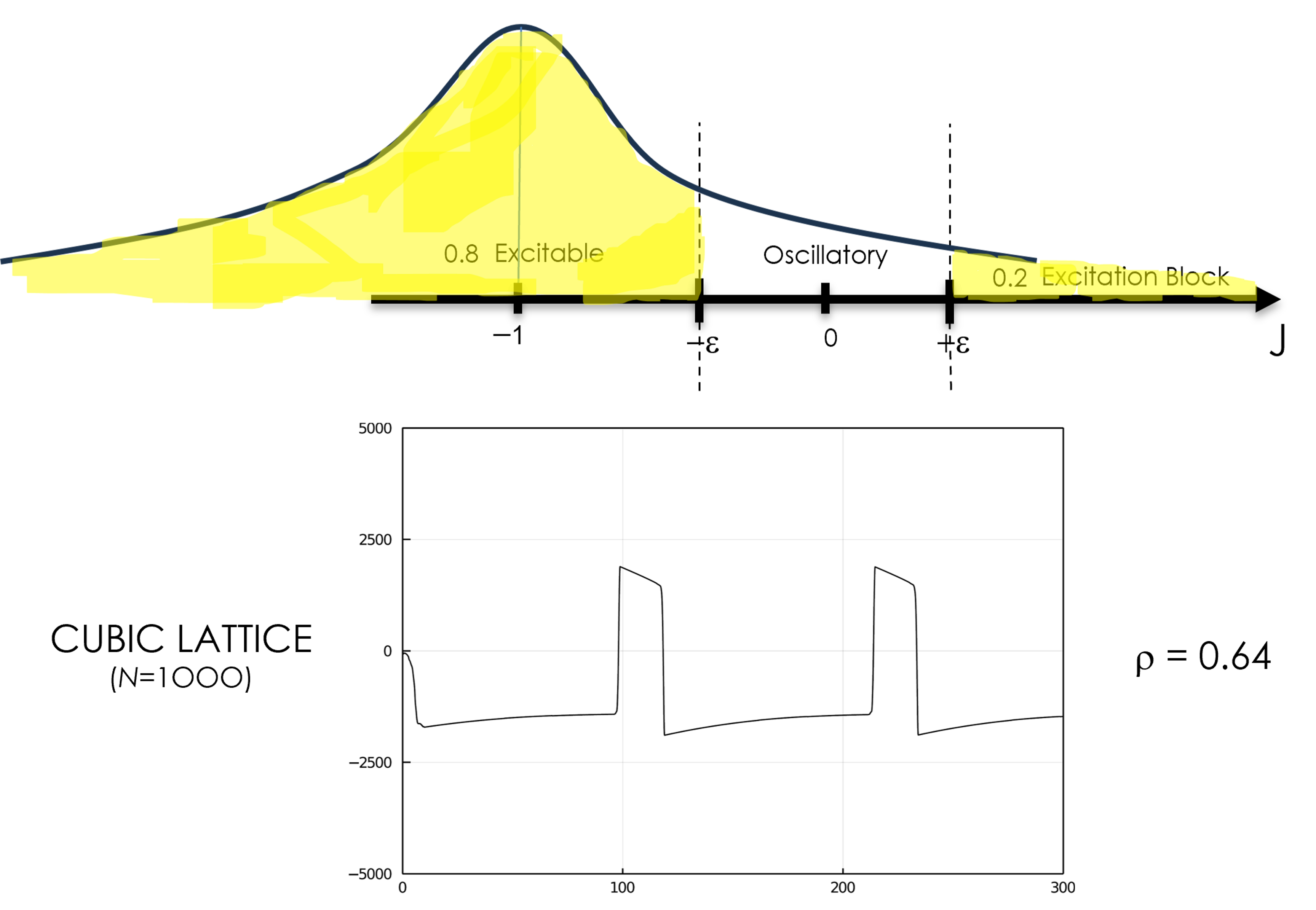}
	\caption{Truncated normal diversity distribution comprising excitable units only, picked from the interval $(-\infty,-\varepsilon)U(+\varepsilon,+\infty)$. The yellow area highlights the portion of the Gaussian that has been used to sample $J_i$ values. The relative symmetry of this distribution produces global network oscillations even in the absence of individually oscillatory units. Both nCOM and the criterion based on oscillator fraction fail to predict this result, however, nCOM correctly indicates that this distribution is more likely to oscillate than that in Fig.~\ref{final1}. The SBS value (SBS = 0.26) is consistent with the observed global oscillations (Table~\ref{tab:asymmetry_summary}).} 
	\label{final2}
\end{figure}

However, if we consider the distribution shown in Fig.~\ref{final2}, where there are no oscillatory units but the excitable ones are now distributed over the range $(-\infty,-\varepsilon)\cup(+\varepsilon,+\infty)$, we observe that the network is again able to exhibit global oscillations, in spite of a diversity distribution that includes $J_i$ values extremely far away from the oscillatory range. 
Rather than from the ratio between oscillatory and excitable units, this behavior can only be inferred from considerations based on the higher symmetry of the distribution in Fig.~\ref{final2} ($\text{nCOM}=24.01$, $\text{SBS} = 0.26$) compared to the one in Fig.~\ref{final1} ($\text{nCOM}=25.15$, $\text{SBS} = 0$). 
Here too, the nCOM parameter fails to predict the global oscillations of the network configuration corresponding to Fig.~\ref{final2}, however, it correctly ranks Fig.~\ref{final1} and Fig.~\ref{final2} in terms of their respective degrees of symmetry and tendencies to give rise to collective oscillations.

To understand the mechanism through which populations of units that are all, or predominantly, in an excitable state can give rise to collective network oscillations, it is instructive to consider the case of a minimal system composed of only two coupled FHN units~\cite{Patriarca2025}. 
Although the FHN model is not a gradient system and does not possess a true potential function, it is still possible to define an effective pseudo-potential for the fast variables under the assumption that the slow variables \(y_i\) are quasi-static over fast timescales~\cite{yamakou2022diversity}. 
For two coupled units, the fast variables obey
\begin{subequations} \label{eq:twoFHN}
\begin{align}
\dot{x}_1 &= a\Bigl[x_1 - \frac{x_1^{3}}{3} + y_1
              + C\bigl(x_2 - x_1\bigr)\Bigr], \label{eq:twoFHN1} \\[2pt]
\dot{x}_2 &= a\Bigl[x_2 - \frac{x_2^{3}}{3} + y_2
              + C\bigl(x_1 - x_2\bigr)\Bigr] \, , \label{eq:twoFHN2}
\end{align}
\end{subequations}
while the slow variables still follow Eq.~\eqref{eq_FN2b}.

Treating $y_1$ and $y_2$ as quasi-constants on the fast timescale
($a\gg1$), one can define an effective pseudo-potential
\begin{equation}
V_{\mathrm{eff}}(x_1, x_2; y_1, y_2)
   = V_1(x_1; y_1) + V_2(x_2; y_2) + \frac{C}{2}(x_1 - x_2)^2,
\label{eq:Veffective}
\end{equation}
where the local terms $V_1$ and $V_2$ are defined as
\begin{equation}
V_i(x_i; y_i) = -\frac{1}{2}x_i^2 + \frac{1}{12}x_i^4 + x_i y_i 
\quad  (i=1,2).
\label{pseudopotential}
\end{equation}
 Each term $V_i(x_i; y_i)$ represents the local pseudo-potential of a single FHN unit at fixed $y_i$, while the quadratic coupling term acts to reduce deviations between the two fast variables, favoring synchronization. 
 When $C = 0$, the landscape consists of two independent tilted wells, and the system relaxes to a fixed point if both units are excitable ($|J_i| > \varepsilon$). 
 However, for moderate coupling, the quadratic term $\tfrac{C}{2}(x_1 - x_2)^2$ deforms the landscape, lowering the barrier between the wells and creating a cyclic valley that can support a limit cycle, even though neither unit would oscillate in isolation.

\begin{figure}[tb]
	%\centering
    \hspace*{-0.5cm}
	\includegraphics[width=0.45\textwidth]{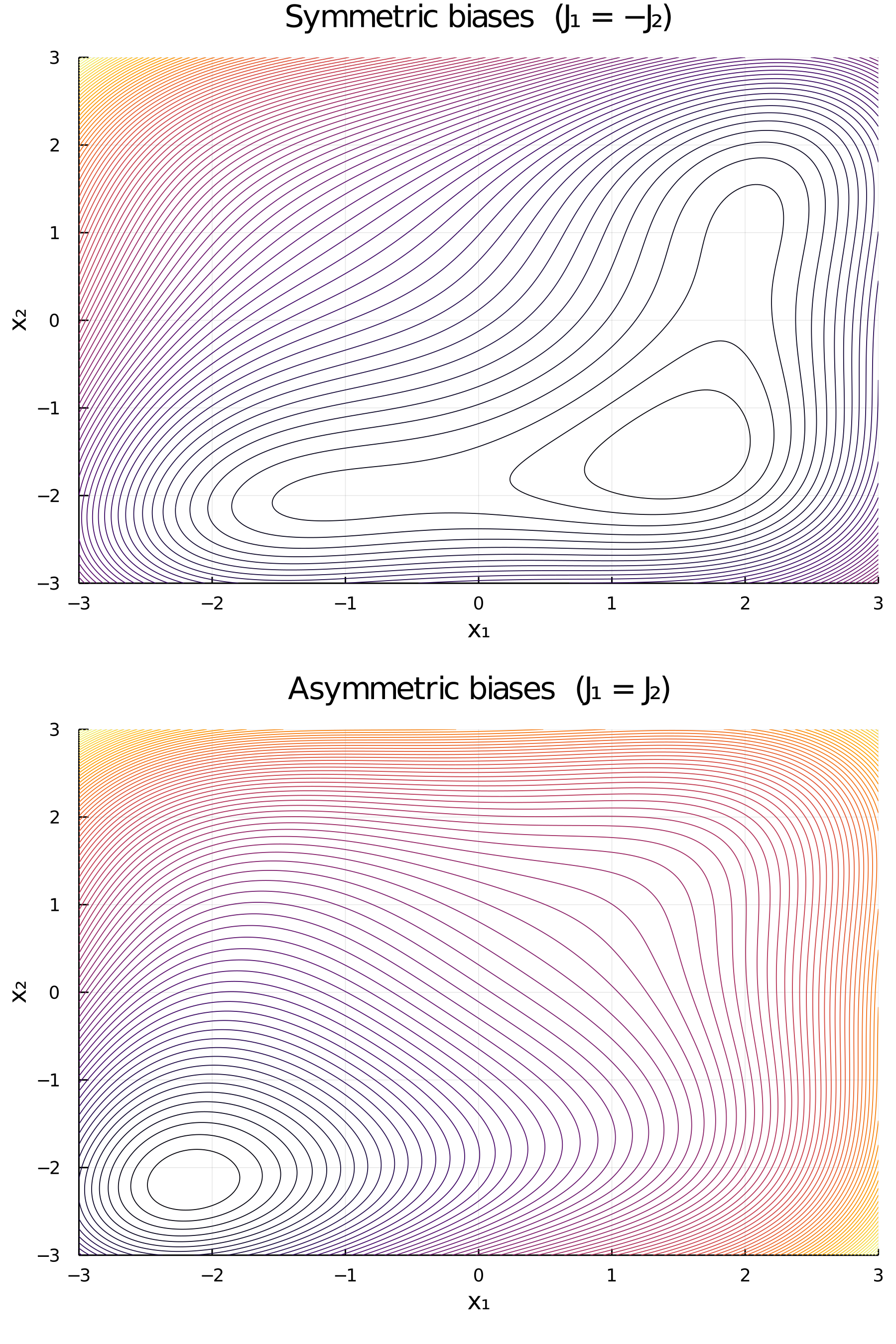}	
	\caption{Contour plots of the effective pseudo-potential defined by Eq.~\eqref{eq:Veffective} in the symmetric (upper panel) and asymmetric (lower panel) cases. In the symmetric case, a double-well-like landscape emerges from the sum of the coupling term with the two local pseudo-potentials $V_1(x_1; y_1)$ and $V_2(x_2; y_2)$, which are tilted in opposite directions due to the different signs of the respective $xy$ terms. In the asymmetric case, the potential landscape collapses into a single minimum as the $xy$ terms have the same sign. Parameter values: $C=0.5$, $\varepsilon=1.0$, $J_2=1.2$ ($J_1=-1.2$ and $J_1=1.2$ in the symmetric and asymmetric case, respectively). The slow variables $y_1$ and $y_2$ are kept constant and set equal to $J_1$ and $J_2$, respectively.} 
    \label{contour}
\end{figure}

Importantly, this collective oscillation can only occur if the respective $J_i$ parameters of the two excitable units are on opposite sides of the oscillatory threshold; that is, they must be positioned symmetrically with respect to the center of the oscillatory interval $(-\varepsilon, +\varepsilon)$. 
This symmetric configuration ensures that the wells of the two local pseudo-potentials are tilted in opposite directions (due to the opposite signs of the $xy$ terms in Eqs.~\eqref{pseudopotential}), forming a double-well-like landscape with a cyclic saddle path that the system can move along (Fig.~\ref{contour}, upper panel). 
If instead the two $J_i$'s are on the same side, the landscape collapses into a single minimum and no oscillatory path can emerge, regardless of coupling (Fig.~\ref{contour}, lower panel).

The two-unit pseudo-potential analysis can be extended to an $N$-unit system by considering a coarse-grained reduction based on the sign of the bias parameters $J_i$. 
Specifically, we partition the population into two subgroups: one composed of units with $J_i > 0$ and the other with $J_i < 0$. 
For each subgroup, we define the average coordinates $(x_+, y_+)$ and $(x_-, y_-)$ by taking the mean of the fast and slow variables across each subpopulation. 
The dynamics of these average states can be approximated by a reduced two-unit system with coupling strength $C$ and effective biases $J_+$ and $J_-$, obtained by taking the average of the $J_i$ values for each subpopulation. 
The effective pseudo-potential then takes the form
\begin{equation}
V_{\mathrm{eff}}(x_+, x_-)
   = V_+(x_+; y_+) + V_-(x_-; y_-) + \frac{C}{2}(x_+ - x_-)^2,
\label{eq:Veffective_clusters}
\end{equation}
where
\begin{equation}
%\begin{align}
V_\pm(x_\pm; y_\pm) = -\frac{1}{2}x_\pm^2 + \frac{1}{12}x_\pm^4 + x_\pm y_\pm  \, ,
%V_-(x_-; y_-) = -\frac{1}{2}x_-^2 + \frac{1}{12}x_-^4 + y_- x_-
%\end{align}
\end{equation}
are the coarse-grained local pseudo-potentials for each cluster.

This reduced formulation, which represents a more accurate approximation when the $J_i$ variability within each subgroup is moderate, captures the essential mechanism by which symmetry in the parameter distribution enables global network oscillations: even when both subgroups are made of excitable units only or comprise a substantial fraction of excitable units, a limit cycle can still emerge provided that their respective average biases lie symmetrically on opposite sides of the oscillatory interval, thus creating a pseudo-potential landscape with opposing tilts and a cyclic valley. 
If instead the subpopulations are biased in the same direction, that is, asymmetrically, the effective landscape collapses into a single minimum, and no global oscillations arise.

\section{Conclusions}
\label{sec_conclusion}

We have demonstrated that the symmetry of the diversity distribution plays a fundamental role in the emergence of global oscillations in networks of coupled excitable units. 
While earlier studies focused on the ratio between oscillatory and excitable elements as the primary driver of collective dynamics, our results reveal that symmetry in the distribution of excitability parameters can act as a more general predictor of oscillatory behavior.

Using both half-normal and truncated normal distributions, we showed that networks with symmetric diversity distributions exhibit robust global oscillations, even when all units are individually excitable, whereas asymmetric distributions tend to suppress collective activity. 
We proposed two complementary metrics, the normalized center of mass (nCOM) and the symmetry balance score (SBS), which quantify different aspects of distribution symmetry and may offer predictive insight into network behavior.

Moreover, an analysis based on effective pseudo-potentials illustrates that symmetry in the bias distribution leads to the formation of a cyclic pathway in the system's potential landscape, enabling limit-cycle oscillations in the collective dynamics. 
In contrast, asymmetry removes this pathway, leading the system to a stable, non-oscillatory state.

These findings deepen our understanding of how structural and statistical properties of heterogeneity shape the collective behavior of excitable systems. 
They suggest that physiological or pathological changes, affecting the balance or symmetry of cellular populations, may have a profound impact on the emergence or loss of coherent dynamics, with potential implications for the modeling of numerous biological systems, such as pancreatic islets, neuronal assemblies, and other excitable cellular networks.
%%%%%%%%%%%%%%%%%%%%%%%%%%%%%%%%%%%%%%%%%%%%%%%%%%%%%%%%%%%%%%%%%%%%%%%%%%%%%%%%%%%%%%%%%%%%%%%%%

%%%%%%%%%%%%%%%%%%%%%%%%%%%%%%%%%%%%%%%%%%%%%%%%%%%%%%%%%%%%%%%%%%%%%%%%%%%%%%%%%%%%%%%%%%%%%%
%\clearpage

\begin{acknowledgments}
S.S., M.P., and E.H. acknowledge support from the Estonian Research Council through grant PRG1059.	M.E.Y. acknowledges support from the Deutsche Forschungsgemeinschaft (DFG, German Research Foundation) via grant YA 764/1-1, Project No.~456989199.
\end{acknowledgments}

%%%%%%%%%%%%%%%%%%%%%%%%%%%%%%%%%%%%%%%%%%%%%%%%%%%%%%%%%%%%%%%%%%%%%%%%%%%%%%%%%%%%%%%%%%%%%%

\bibliography{cellulebeta22dic2020.bib}

\end{document}